Creative Education, 2025, 16(7), *-*
https://www.scirp.org/journal/ce
ISSN Online: 2151-4771
ISSN Print: 2151-4755# Pictorial and Documentary Guide for Research, Teaching, and Education through Astronomy, Physics, and Mathematics Pursued under the Umbrella of the United Nations (1974-2024)

Hans J. Haubold[1,2], Arak M. Mathai[3]

[1]Outer Space Affairs Division, United Nations, New York, USA
[2]Office for Outer Space Affairs, United Nations, Vienna, Austria
[3]Department of Mathematics and Statistics, McGill University, Montreal, Canada
Email: hans.haubold@gmail.com, directorcms458@gmail.comHow to cite this paper: Haubold, H. J., & Mathai, A. M. (2025). Paper Title. *Creative Education, 16,* **-**.
https://doi.org/10.4236/***.2025.*****

Received: **** **, ***
Accepted: **** **, ***
Published: **** **, ***Copyright © 2025 by author(s) and Scientific Research Publishing Inc.
This work is licensed under the Creative Commons Attribution International License (CC BY 4.0).
http://creativecommons.org/licenses/by/4.0/

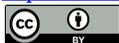## Abstract

This paper was prepared for Open-Access-only publication as a guide reporting on Education (all aspects of space science and technology), Teaching (remote sensing and GIS, satellite meteorology and global climate, satellite communication, space and atmospheric sciences, global navigation satellite systems), and Research (solar neutrino problem, formation of structure in the Universe) in astronomy (solar physics, cosmology), physics (nuclear physics, neutrino physics), and mathematics (fractional calculus, special functions of mathematical physics) exercised over a period of 50 years (1974-2024). In this period, more than twenty workshops were held and seven regional centres for space science and technology education were established over all regions of the world: Asia and the Pacific, Latin America and the Caribbean, Africa, Western Asia, and Europe. This effort was undertaken in cooperation of ESA, NASA, JAXA, and 193 member states of the United Nations under the auspices of the UN, also supported by the Committee on Space Research (COSPAR) and the International Astronomical Union (IAU). The paper provides access to most of the documents in the six official languages of the United Nations (Arabic, Chinese, English, French, Russian and Spanish), proceedings, and published papers and books focusing on education, teaching, and research (listed in Google Scholar and Research Gate).

## Keywords

Education, Teaching, Research, Astronomy, Physics, Mathematics, International Cooperation, Regional Cooperation, United Nations (UN),DOI: 10.4236/***.2025.*****   **** **, 2025   1   Creative Education

European Space Agency (ESA), National Aeronautics and Space Administration (NASA), Japan Aerospace Exploration Agency (JAXA)

## 1. Introduction: A Holistic Approach to Einstein and the United Nations

In 1700, Gottfried Wilhelm Leibniz initiated the establishment of a Society of Sciences, which was to evolve into the Academy of Sciences with its motto "heoria cum praxi" that much later incorporated the Astrophysical Observatory Potsdam (Germany, **Figure 1**) and the Albert Einstein Laboratory of Theoretical Physics Caputh (Germany, **Figure 2**). The latter was directed by Hans-Juergen Treder and located in Einstein's Summerhouse Caputh (Strauch, 2015).

In 1974, it was Treder who advised the authors to take up the solar neutrino problem for prospective research. At that time, only solar neutrino data from the Homestake experiment (USA) were available and a prospective solution of the problem was expected by new solar, nuclear, or neutrino physics. Treder reminded the authors of the history of the development of the theory of quantum radiation as pursued by Ludwig Boltzmann, Max Planck, Albert Einstein, Erwin Schrödinger, and Werner Heisenberg and encouraged speculations that similar physics might be discovered for neutrino radiation. A closer study of the Proceedings of the first Solvay Conference held in 1911 (Nernst, 1914) provided advice to look closer to Henri Poincaré's comment on new mathematics and Einstein's comment on the use of Boltzmann-Gibb's entropy in the light of the solar neutrino problem.

The idea of fractional calculus can be traced for the first time to a letter exchange between Gottfried Wilhelm Leibniz and the Marquis de l'Hôpital in 1695, in which Leibniz expressed the idea of the existence of derivatives of non-integer orders (**Figure 3**, **Figure 4** and 30). One of the most interesting applications of fractional calculus can be considered the entropy production paradox for fractional (anomalous) diffusion that is supposed to bring more light on understanding the disperse diffusion and wave propagation. It was Poincaré who reminded the Solvay Conference participants that the development of quantum mechanics may need new mathematics (different from methods with differential calculus). At the same conference, it was Einstein who reminded participants that "neither Herr Boltzmann nor Herr Planck gave a definition of W (probability in Boltzmann-Gibbs entropy). They put formally W = number of complexions of the state under consideration". For the authors of the current paper, this was encouragement to apply anomalous diffusion entropy analysis to the solar neutrino data from the SuperKamiokande experiment.

Considering the specific circumstances in writing this paper, the authors want to have a side-look at the relation between Einstein and the United Nations (see below) shown in **Figure 5**. The colouring with the green-yellowish shades versus reddish tones is meant to distinguish sciences from humanities, coinciding with the left and right sides of the giant component. These sides are linked by Sir James Chadwick, who won the 1935 Nobel Prize in Physics for discovering the neutron and who also became a scientific advisor to the United Nations. The science side, making visible researchers like Albert Einstein and Max Planck and founders of modern physics from the Curies to Enrico Fermi and Eugene Wigner or György Hevesy.

On the humanities side, one can see some well-known individuals. There are two central laureate organizations that strike the eye: the European Union and the United Nations, both awarded Nobel Peace Prizes. Notable individuals include prominent politicians, such as Barack Obama or Henry Kissinger, the human rights activist Nelson Mandela, and the economist Milton Friedman. Much more details on the development of and in-depth information in **Figure 5** are highlighted by the author, Milan Janosov, at https://arxiv.org/pdf/2309.15610

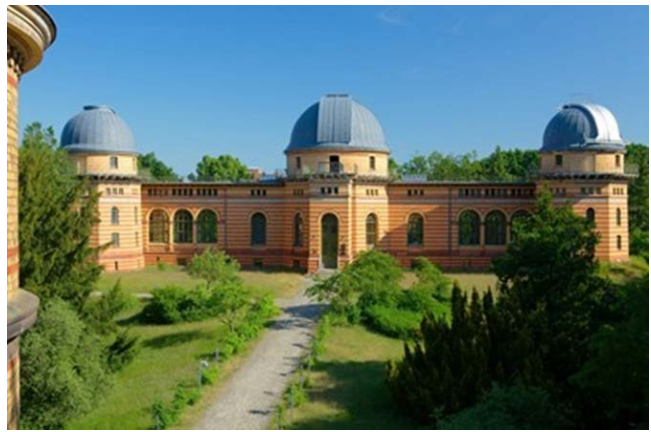

Figure 1. Astrophysical observatory Potsdam, Potsdam, Germany.

https://researchfeatures.com/nonadditive-entropies-generalising-boltzmanns-approach-thermodynamics/

https://researchfeatures.com/perplexing-solar-investigation-experimenting-neutrinos/

https://researchoutreach.org/articles/exploring-neutrinos-entanglement-entropy-fractional-calculus/

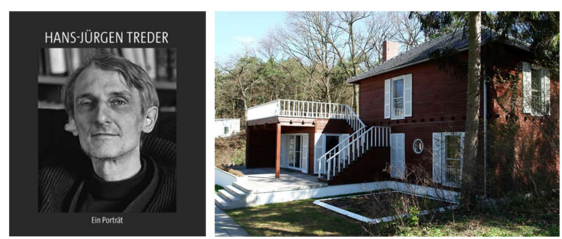

Figure 2. Hans-Jürgen Treder, former Director of the Astrophysical Observatory Potsdam, Germany, (left) and director of the Albert Einstein Laboratory for Theoretical Physics, Caputh, Germany (right).

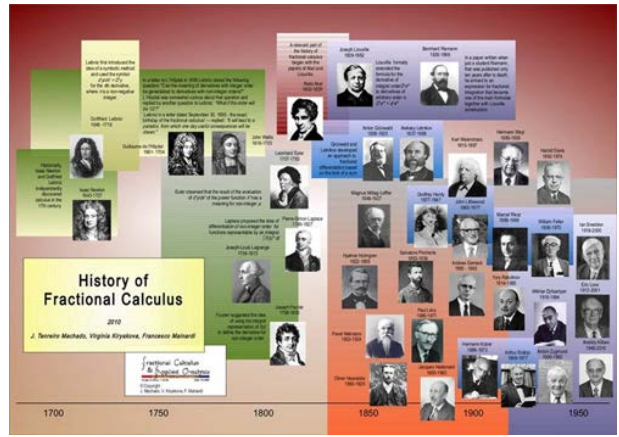

**Figure 3.** The timeline of fractional calculus during the period 1695-1970 in J. A. Tenreiro Machado, V. Kiryakova, F. Mainardi, Fractional Calculus & Applied Analysis, 13 (2010) 447-454.

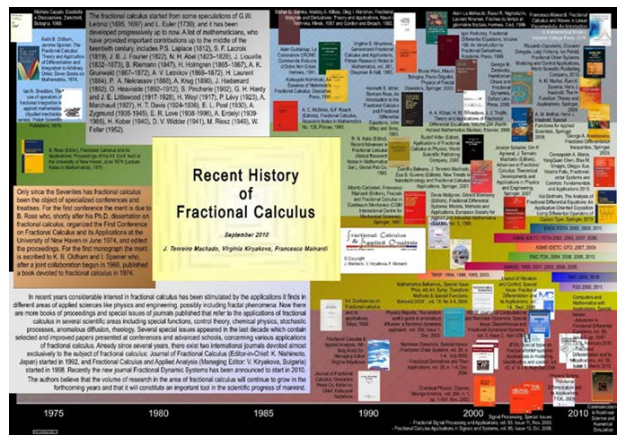

**Figure 4.** The timeline of fractional calculus during the period 1966-2010 in J.A. Tenreiro Machado, V. Kiryakova, F. Mainardi, Fractional Calculus & Applied Analysis 13 (2010) 329-334.

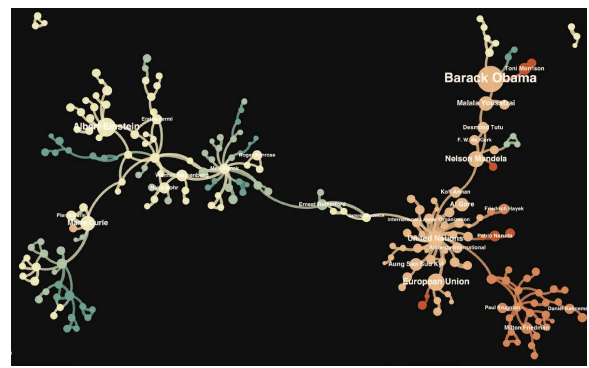

**Figure 5.** Nobel Network. The network of Nobel laureates with at least one connection, based on the cross-references between their Wikipedia pages. Each node corresponds to a laurate, edge widths measure the number of cross references, and node size is proportional to the total view count of their Wiki pages. Colour encodes the disciplines they were awarded (in the case of multiple different awards, a colour was picked at random from the awarded disciplines). Nodes with the highest view counts are labelled.

Research in Nuclear and Neutrino Astrophysics, 1974-2024 (Haubold & Mathai, 1997; Reines, 1972), Pursued at the Astrophysical Observatory Potsdam, Potsdam-Telegrafenberg, Germany.

The solar neutrino problem was named the large discrepancy between the flux of solar electron neutrinos as predicted from the solar standard model and as measured on Earth in Raymond Davis's Homestake experiment in the 1970s. Around 2002, the problem was considered solved after checking all aspects of the standard solar model and the standard model of elementary particles. Physicists were aware that a mechanism, discussed in 1957 by Bruno Pontecorvo, could explain the deficit in electron neutrinos. The authors contributed to the international efforts to solve the solar neutrino problem by attempting to derive analytic models of the internal structure of the Sun and analytic closed form representations of the reaction probability for nuclear reactions proceeding in the core of the Sun. In such attempts, it was discovered that the theory of special functions of mathematical physics is very helpful. The Mittag-Leffler function can be considered the queen function of the fractional calculus, and the Fox H-function has been shown to be central to fractional quantum mechanics (**Figure 6**).

As initiated by Hans-Juergen Treder, the books in **Figure 6** explore the fundamental workings of the Sun, unravelling the mysteries of solar structure, neutrino emission, and the profound interplay between solar physics, nuclear reactions, and neutrino properties. Diving into the intricate world of nuclear and neutrino astrophysics with these books, beginning with the renowned solar neutrino problem, the books recount the historical journey that commenced in 1968 with experiments aiming to detect solar neutrinos. These pioneering efforts unveiled a significant discordance between predicted solar neutrino fluxes and actual measurements—a puzzle that persisted until its resolution in 2002. The authors, immersed in this quest since 1974, undertook a multifaceted approach, refining theories in solar physics, nuclear physics, and neutrino science. Central to their investigation were analytic mathematical methods, particularly the application of special functions of mathematical physics. The narrative unfolds with a meticulous exploration of solar model construction, presenting simplified models that capture the Sun's complexity. Notably, the authors provide closed-form presentations of nuclear reaction rates, a critical cornerstone in understanding stellar energy generation and neutrino production. Moreover, the books showcase advanced techniques for analysing solar neutrino data, offering invaluable insights into the intricate dance of particles within the solar core. For researchers captivated by the enigmatic realm of astrophysical phenomena, the books serve as an indispensable guide. It encapsulates decades of rigorous research and intellectual pursuit, presenting a comprehensive synthesis of theoretical advancements and empirical findings in nuclear and neutrino astrophysics. Whether delving into solar dynamics, nuclear processes, or neutrino behaviour, readers will find themselves enriched by the profound contributions and analytical acumen. The books are also providing an insight of cooperation with leading scientists in this field of research like the Nobel prize winners W.A. Fowler and R. Davis Jr.

H.-J. Treder: https://adsabs.harvard.edu/full/1974an....295..169t

E. Gerth: https://articles.adsabs.harvard.edu/pdf/1983AN....304..299H

R. Davis, Jr. (Nobel Price 2002):

https://www.nobelprize.org/prizes/physics/2002/davis/facts/

The research topic "fractional quantum diffusion and solar neutrino entanglement entropy production" was further pursued in the workshops (see below) organized under the umbrella of the United Nations.

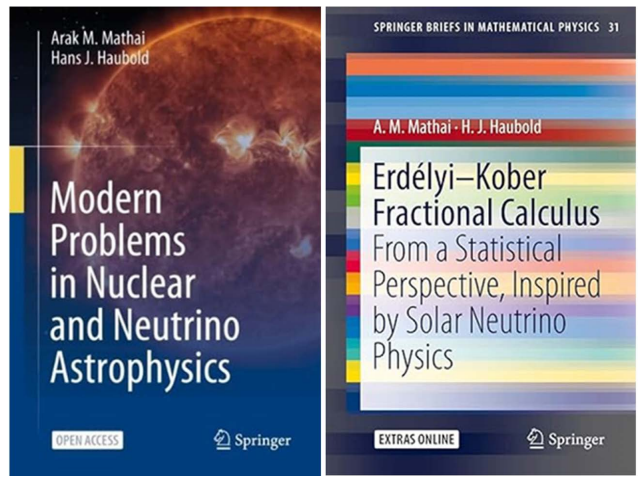

**Figure 6.** Report summarizing research results of the period 1974 to 1988 (Meijer's G-function, left hand) and report summarizing research results of the period 1988 to 2025 (Fox' H-function, right hand).

A.M. Mathai:

https://wol-prod-cdn.literatumonline.com/doi/10.1002/mana.19710480110

R.W. John: 1978AN....299..225H (harvard.edu)

W.A. Fowler (Nobel Price 1983):

https://www.nobelprize.org/prizes/physics/1983/fowler/biographical/

**Research on the Early Evolution of the Universe and Formation of Structure, 1984-1991 (**Haubold **& Mathai, 1998;** Alpher **& Herman, 2001;** Gottloeber **et al., 1990), Conducted at the Astronomical Observatory, Potsdam-Babelsberg, Germany.**

In 1986, a working group on the formation of structure in the Universe was established (**Figure 7**), involving the authors, to extend individual research topics covering a wide range from relativistic astrophysics to gravitation. Alpher and Herman worked out the detailed evolution of the early Universe and did show in collaboration with Follin a quantitative depiction of the synthesis of light elements in the early Universe and predicted that there should be a cosmic neutrino background bath at a slightly lower temperature than that of cosmic background radiation. They emphasized already in the early 1950s that this neutrino background and solar neutrino radiation remain to be discovered. So do the even more elusive gravitational radiation background the three authors predicted in their research publications.

In the book (**Figure 8**), significant results are presented regarding the formation of structure in the universe and the key conclusions are put into the context of future research directions. Attention is given to the failure of hot big-bang cosmology to account for observable cosmological structures despite the solid foundations of the big-bang descriptions early universe evolution. The problems of classical cosmology are listed, and descriptions are given of the major tenets of the inflationary scenario and the linear vs nonlinear approaches to gravitational instability. Luminous matter is interpreted based on absorption models and the clustering in quasar absorption lines, and the creation of chemical elements is discussed in terms of nucleosynthesis and physical evidence for nucleosynthesis. It is concluded that a close connection exists between the problems of the very early universe solved in the inflationary scenario and the problems of structure formation and chemical evolution. The book represents a sample research programme on topics of the early evolution of the Universe and formation of structure.

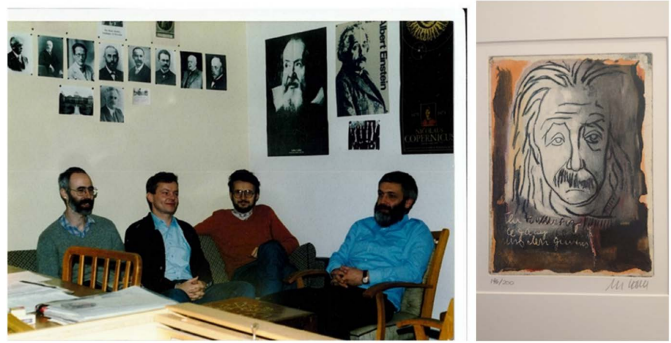

**Figure 7.** Left: J.-P. Muecket, HJH, V. Mueller, and S. Gottloeber, working group on formation of structure in the Universe, in the office at the Institute for Astrophysics, Potsdam.

Right: Armin Mueller-Stahl (https://de.wikipedia.org/wiki/Armin_Mueller-Stahl), Colour etching The Enduring Legacy of a Modern Genius: Einstein, 2014.

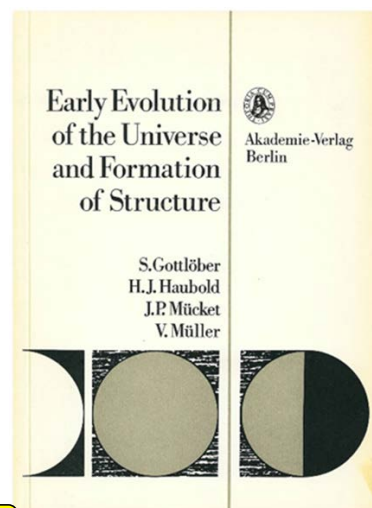

**Figure 8.** Early Evolution of the Universe and Formation of Structure.
https://www.degruyter.com/document/isbn/9783112754146/html?lang=en

A.M. Mathai:

https://pubs.aip.org/aip/jmp/articl-bstract/29/9/2069/228029/Gravitational-in-stability-in-a-multicomponent?redirectedFrom=fulltext

**Educational Celebration of the Einstein Centenary 1979 in Berlin, Germany, and the Michelson Colloquium 1981 in Potsdam, Germany (Haubold, 2024; Treder & Rompe, 1982).**

The celebrations of Albert Einstein's 100th birthday in Berlin 1979 and of Albert A. Michelson's 100th anniversary of his ether drift experiment at the Astrophysical Observatory Potsdam (AOP) 1981 have been prepared by setting up a comprehensive exhibition at the AOP. This was designed as an education, teaching, and research event to make visible all still existing instruments and publications for national and international visitors. The chief guest and lecturer at the Michelson Colloquium in Potsdam was the youngest daughter of Albert A. Michelson from New York (USA), Mrs. Dorothy Michelson Livingston (**Figure 9**). An attraction of the exhibition was the cellar in the main building of the AOP where Michelson performed his experiment in 1881 (**Figure 9**). Original documents were exhibited that have shown the strong support of the then director of the AOP and the famous physicist Hermann von Helmholtz of the University of Berlin.

E. Yasui: https://link.springer.com/article/10.1007/bf00412263

L. Pyenson:

https://pubs.aip.org/aip/acp/article-abstract/179/1/42/747229/Michelson-s-first-ether-drift-experiment-in-Berlin?redirectedFrom=fulltext

A.A. Michelson (Nobel Price 1907):

https://www.nobelprize.org/prizes/physics/1907/michelson/biographical/

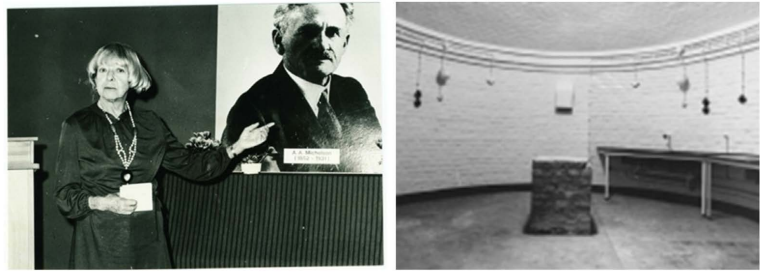

Dorothy Michelson Livingston, youngest daughter of Abert A. Michelson, opening the Michelson Colloquium in Potsdam, Germany, 27-30 April 1981.

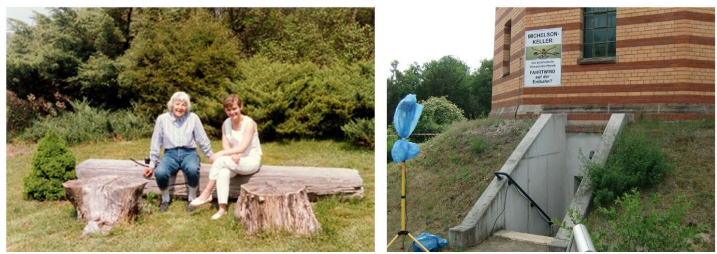

**Figure 9.** 1989 Dorothy Michelson Livingston with BH on the Michelson Livingston premises in Wainscott, New York (left) and the entrance of the Michelson cellar (Astrophysical Observatory Potsdam) (right).

Educational Celebration of the Michelson Experiment with a View to the Hudson River School, New York, USA, 1974-2024 (Michelson Livingston, 1973).

The Einstein centenary 1979 and the Michelson Colloquium 1981 lead to a co-operation between Mrs. Michelson Livingston and one of the authors, starting in 1978 in Potsdam until 1994 in New York, while one of the authors was working at the United Nations in New York city (**Figure 10**). In this time, documents were checked which Mrs. Michelson Livingston collected while writing the biography of her father. Some of these documents like the letters she exchanged with Max Born and Helen Dukas (Einstein's livelong secretary) concerning the importance of the Michelson experiment for Einstein's relativity theory were unearthed and prepared for later publication. In this connection, also Michelson's interest for landscape painting and the role of the Hudson River School in the development of art in the United States were discussed. It came to light that she had already visited the AOP in the 1960$^s$ and was made aware of Michelon's activities at AOP and the University of Berlin by Johann Wempe, then the director of AOP.

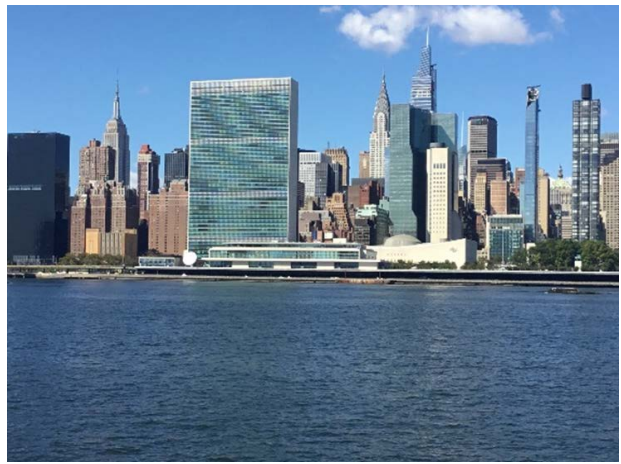

Headquarter of the United Nations at East River in New York.

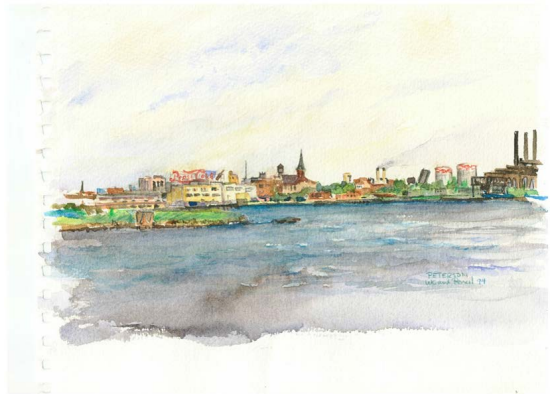

**Figure 10.** Jerald Peterson (http://www.jeraldpeterson.com/), Watercolour view of the landscape of Queens across the East River from HJH's office 32$^{nd}$ floor United Nations Building, New York.

Albert A. Michelson had many interests outside of the realm of physics and science, including music, art, billiards, chess, and tennis. His interest in the violin began when he was a child, grew during his years at the Naval Academy, and eventually led to musical composition. Michelson expressed his interest in art through sketching and painting watercolours. Many of Michelson's paintings were completed in his later years and during his retirement, consisting mostly of watercolours of California landscapes. Some of Michelson's watercolours were exhibited at the Pasadena Art Institute in 1931, shortly before his death. Michelson was also an artist. In 1928, he had a one-man exhibition at the Renaissance Society of the University of Chicago and exhibited two watercolours (Antofagasta, Chile and Vigo Harbor, Spain) at the Art Institute of Chicago's Eighth Annual Exhibition of Watercolours by American Artists. At one of these exhibits, "one woman told Michelson he should never have given up art for science. 'I did not have to choose,' he answered, 'because for me they are inseparable'". Indeed, 25 years previously, in his book Light Waves and Their Uses (1903), Michelson had written: And painting lessons each Sunday from Rudolph Weisenborn at the Chicago Academy of Fine Arts and amused himself by drawing caricatures of acquaintances. After he remarried, he built a house in Chicago in 1923 with a conservatory in which his wife grew flowers and both sat and drew or painted, a habit they took outdoors in their later years in Pasadena, California. As a supplement communication during the Michelson Colloquium, Mrs. Dorothy Michelson Livingston made a lively presentation with photography and paintings on her father as an artist in science and technology.

https://researchfeatures.com/dorothy-michelson-livingston-personal-recollection/ and

https://researchfeatures.com/art-science-connecting-motives-research-landscape-painting-19th-century/

**Research Material and Lecture Courses at the A.M. Mathai Centre for Mathematical and Statistical Sciences, Kerala, India, 1982-2011** (Haubold, 2020; Mathai, 2011).

In 1982, the cooperation between the two authors started with the question on how to cover research, teaching, and education in astronomy, physics, and mathematics in an established centre for mathematics and space sciences. Preparations were already on-going in the National Centre for Mathematical and Statistical Sciences located in India. Until 2011 the centre established an infrastructure for fully operating education, teaching, and research in mathematics, statistics, and related scientific disciplines with continuing international cooperation (**Figure 11**).

In Haubold, 2020, a brief history of the Centre for Mathematical and Statistical Sciences, Kerala, India, is given and an overview of Mathai's research and education programs in the following topics is outlined: Fractional Calculus; Functions of Matrix Argument—M-transforms, M-convolutions; Krätzel integrals; Pathway Models; Geometrical Probabilities; Astrophysics—reaction rate theory, solar neutrinos; Special Functions—G and H-functions; Multivariate Analysis; Algorithms for Non-linear Least Squares; Characterizations—characterizations of densities, information measure, axiomatic definitions, pseudo analytic functions of matrix argument and characterization of the normal probability law; Mathai's Entropy—entropy optimization; Analysis of Variance; Population Problems and Social Sciences; Quadratic and Bilinear Forms; Linear Algebra; Probability and Statistics, Multivariate Statistical Analysis in the Real and Complex Domains (**Figure 12**).

https://www.growkudos.com/projects/a-m-mathai-centre-for-mathematical-and-statistical-sciences-nurturing-the-love-for-mathematics

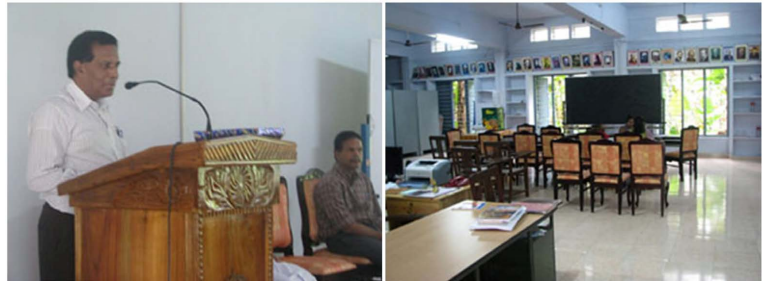

A.M. Mathai, Department of Mathematics and Statistics, McGill University, Montreal, Canada, and Director of the Centre for Mathematical and Statistical Sciences, Kerala, India.

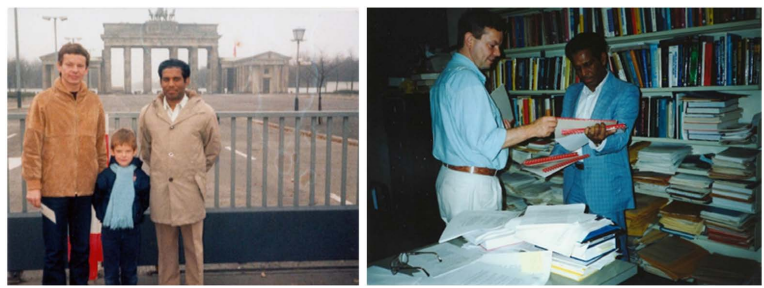

**Figure 11.** Left: HJH, Haubold junior, A.M. Mathai, in front of the Brandenburg Gate Berlin in 1984. Right: HJH and A.M. Mathai in his office at McGill University Montreal in 1992.

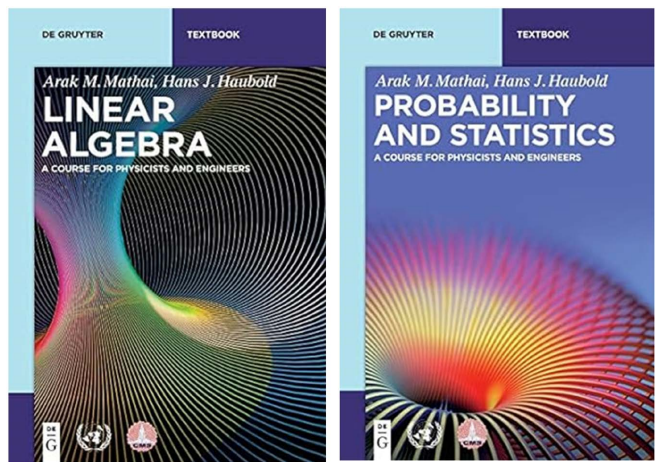

Linear Algebra and Probability and Statistics for teaching basic courses in the following five aeras of space science and technology.

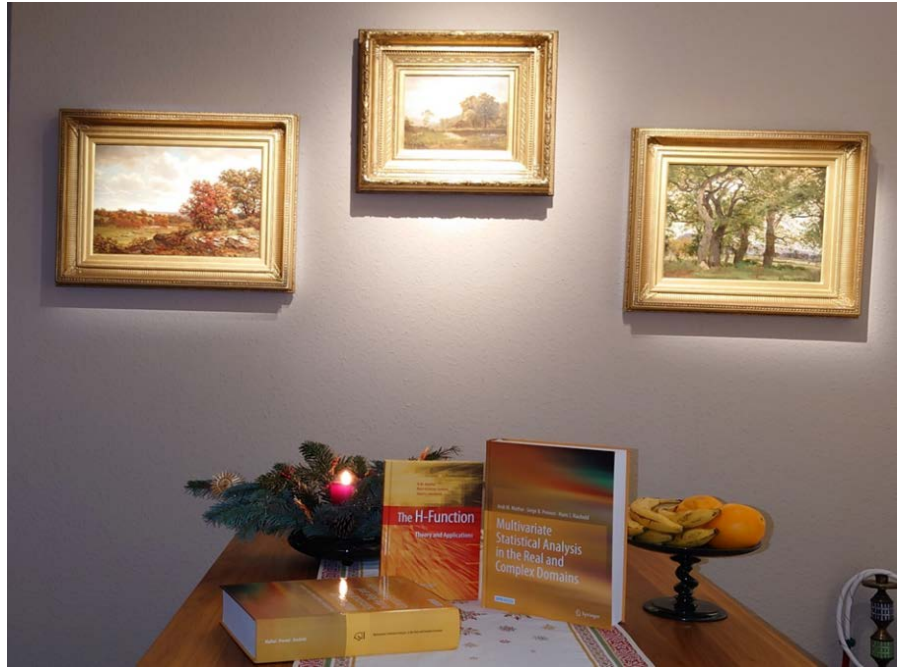

**Figure 12.** Three paintings from Brown, Bierstadt, and Whittredge (from left to right) used to demonstrate the link between landscape painting and science. Also shown are the monographs on Fox's H-function and Multivariate Statistical Analysis in the Real and Complex Domains.

**Teaching Curricula Prepared for Regional Centres for Space Science and Technology Education, New York, USA, 1988-1993 (Pyenson et al., 2019; Capacity building, 2008).**

1988 one of the authors was relocated to the Outer Space Affairs Division of the United Nations in New York were member states had initiated the project of establishing regional centres for space science and technology education, covering five focal scientific and technological topics. For this project the expertise from CMS (Mathai, 2011) was utilized for the regional centres and the development of education curricula proceeded in expert meetings hosted by UK, Spain, and Italy (Capacity building, 2008). From among the teaching material published by CMS, three graduate books on algebra and probability and statistics were made available for teaching at the regional centres (**Figure 12**).

Based on resolutions of the United Nations General Assembly, UN-affiliated Regional Centres for Space Science and Technology Education were established in China, India, Morocco, Nigeria, Brazil, Mexico, and Jordan. Simultaneously, education curricula at the university level were developed for the core disciplines of remote sensing and the geographic information system, satellite communications, satellite meteorology and global climate, space and atmospheric science, and global navigation satellite systems (**Figure 13**). In 2017, these education curricula were supplemented by the publication of two open access books covering full university courses in linear algebra and probability and statistics. The courses have been taught over several decades at McGill University Montreal, Canada, and at the Centre for Mathematical and Statistical Sciences, Kerala, India, by A.M. Mathai.

A.M. Mathai Centre for Mathematical and Statistical Sciences and, subsequently,

Six UN-affiliated Regional Centres for Space Science and Technology Education (affiliated to the United Nations)

https://researchfeatures.com/promoting-global-education-teaching-research-space-science/

https://www.unoosa.org/pdf/publications/st_space_41E.pdf

1) Africa, Nigeria (ARCSSTE-E)

2) Africa, Morocco (CRASTE-LF)

3) Latin America & Caribbean, Mexico & Brazil (CRECTEALC)

4) Asia/Pacific, India (CSSTEAP)

5) West Asia, Jordan (RCSSTEWA)

6) Asia/Pacific, China (RCSSTEAP)

1. SATELLITE METEOROLOGY AND GLOBAL CLIMATE

https://www.unoosa.org/documents/pdf/psa/reg_centres/curricula/Satellite_Metorology_and_Global_Climate.pdf

2. SATELLITE COMMUNICATIONS

https://www.unoosa.org/documents/pdf/psa/reg_centres/curricula/Satellite_Communications.pdf

3. SPACE AND ATMOSPHERIC SCIENCE

https://www.unoosa.org/documents/pdf/psa/reg_centres/curricula/Space_and_Atmospheric_Science.pdf

4. REMOTE SENSING AND GEOGRAPHIC INFORMATION SYSTEM

https://www.unoosa.org/documents/pdf/psa/reg_centres/curricula/Remote_Sensing_and_GIS.pdf

5. GLOBAL NAVIGATION SATELLITE SYSTEMS

https://www.unoosa.org/pdf/icg/2013/Ed_GNSS_eBook.pdf

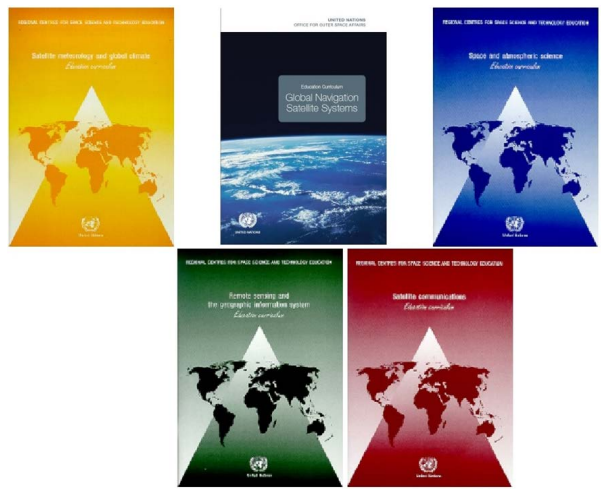

**Figure 13.** Education Curricula space science and technology at the University level.

Education Workshops in the Fields of Astronomy, Physics, and Mathematics, Prepared at the United Nations, New York, USA, 1988-2012 (Mathai & Haubold, 2018).

Additionally, to the establishment of regional centres for space science and technology education, UN member states recommended the holding of consecutive annual workshops on basic space science to identify topics at the regional and international level that may be pursued in universities and research institutes for the benefit of academia in member states. Four workshops on basic space science were held, one in each of the four regions on Earth.

The UN General Assembly documents available through Open Access in this section contain an overview and summary on the achievements of the basic space science workshops in terms of donated and provided planetariums, astronomical instruments, and space weather instruments, particularly operating in developing nations. These instruments have been made available to respective host countries, particularly developing nations, through the series of twenty basic space science workshops, organized by the United Nations Programme on Space Applications 1991-2012. Organized by the United Nations, the European Space Agency (ESA), the National Aeronautics and Space Administration (NASA) of the United States of America, and the Japan Aerospace Exploration Agency (JAXA), the basic space science workshops were organized as a series of workshops that focused on basic space science (1991-2004), the International Heliophysical Year 2007 (2005-2009), and the International Space Weather Initiative (2010-2012) proposed by the UN Committee on the Peaceful Uses of Outer Space on the basis of discussions of its Scientific and Technical Subcommittee, as reflected in the reports of the Subcommittee. Workshops on the International Space Weather Initiative in the series were hosted by the Government of Egypt in 2010 (see A/AC.105/994), the Government of Nigeria in 2011, and the Government of Ecuador in 2012 (see A/AC.105/1030). Workshops on the International Heliophysical Year 2007 were hosted by the United Arab Emirates in 2005 (see A/AC.105/856), India in 2006 (see A/AC.105/882), Japan in 2007 (see A/AC.105/902), Bulgaria in 2008 (see A/AC.105/919) and the Republic of Korea in 2009 (see A/AC.105/964). Workshops on basic space science were hosted by the Governments of India (see A/AC.105/489), Costa Rica and Colombia (see A/AC.105/530), Nigeria (see A/AC.105/560/Add.1), Egypt (see A/AC.105/580), Sri Lanka (see A/AC.105/640), Germany (see A/AC.105/657), Honduras (see A/AC.105/682), Jordan (see A/AC.105/723), France (see A/AC.105/742), Mauritius (see A/AC.105/766), Argentina (see A/AC.105/784) and China (see A/AC.105/829). All workshops were co-organized by the International Astronomical Union (IAU) and the Committee on Space Research (COSPAR). Below the reports on each of the 20 workshops are Open Accessible in the six official languages of the United Nations.

Four Workshops **Basic Space Science (**Wamsteker et al., 2004**).**

01_India_1991 (report not available online) (**Figure 14**)

REPORT ON THE UNITED NATIONS/EUROPEAN SPACE AGENCY WORKSHOP ON BASIC SPACE SCIENCE FOR DEVELOPING COUNTRIES, ORGANIZED IN COOPERATION WITH THE GOVERNMENT OF INDIA AND HOSTED BY THE INDIAN SPACE RESEARCH ORGANIZATION (Bangalore, India, 30 April-3 May 1991)

Proceedings: https://pubs.aip.org/aip/acp/issue/245/1

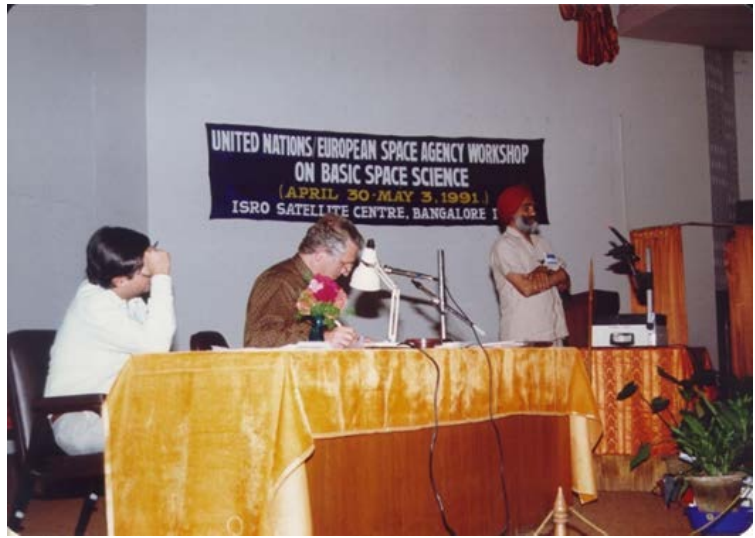

Figure 14. S. Chakravarty (India), W. Wamsteker (ESA), and a participant opening the 1991 workshop in India.

02_Columbia and Costa Rica_1992 (report not available online).

REPORT ON THE SECOND UNITED NATIONS/EUROPEAN SPACE AGENCY WORKSHOP ON BASIC SPACE SCIENCE FOR DEVELOPING COUNTRIES, CO-SPONSORED BY THE EUROPEAN SPACE AGENCY AND THE PLANETARY SOCIETY, ORGANIZED IN COOPERATION WITH THE GOVERNMENTS OF COSTA RICA AND COLOMBIA AND HOSTED BY THE UNIVERSITY OF COSTA RICA, THE INTERNATIONAL CENTRE FOR PHYSICS AND THE UNIVERSITY OF THE ANDES (San Jose, Costa Rica, and Santa Fe de Bogota, Colombia, 2-13 November 1992)

Proceedings:

https://link.springer.com/journal/10509/volumes-and-issues/214-1

Proceedings:

https://link.springer.com/journal/11038/volumes-and-issues/63-2

03_Nigeria_1993 (report not available online)

REPORT ON THE THIRD UNITED NATIONS/EUROPEAN SPACE AGENCY WORKSHOP ON BASIC SPACE SCIENCE FOR DEVELOPING COUNTRIES, ORGANIZED IN COOPERATION WITH THE GOVERNMENT OF NIGERIA AND HOSTED BY THE UNIVERSITY OF NIGERIA, NSUKKA, AND THE OBAFEMI AWOLOWO UNIVERSITY, ILE-IFE (Lagos, Nigeria, 18-22 October 1993).

Proceedings: https://pubs.aip.org/aip/acp/issue/320/1

04_Egypt_1994 (report not available online)

REPORT ON THE FOURTH UNITED NATIONS/EUROPEAN SPACE AGENCY WORKSHOP ON BASIC SPACE SCIENCE, HOSTED BY THE GOVERNMENT OF EGYPT (Cairo, Egypt, 27 June-1 July 1994).

Proceedings:

https://link.springer.com/journal/10509/volumes-and-issues/228-1

Proceedings:

https://link.springer.com/journal/11038/volumes-and-issues/70-1

**Education Workshops in the Fields of Astronomy, Physics, and Mathematics, prepared at the United Nations, Vienna, Austria, 1993-2012 (Continued) (Figure 15).**

Subsequent to the four regional workshops, eight annual workshops focusing on specific topics in space science and technology, as recommended by experts in the four regional workshops 1991-1994, were held in the period 1995-2004. Pursuant to the UN declaration of 2007 to be the International Heliophysical Year (IHY2007), five workshops to prepare and to follow up the IHY 2007, were held from 2005-2009.

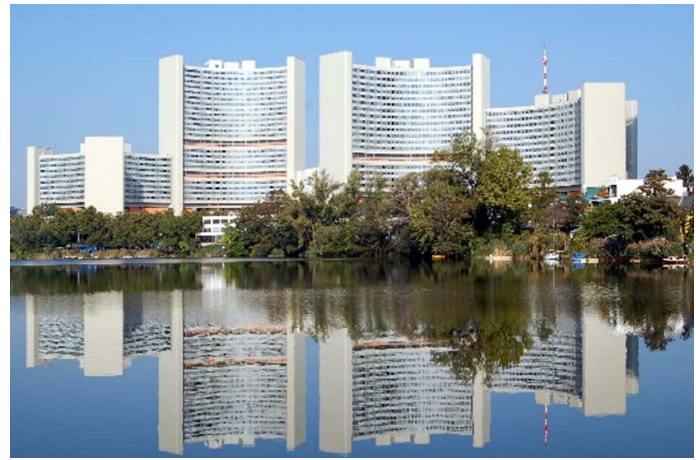

Headquarter of the United Nations at Danube River in Vienna.

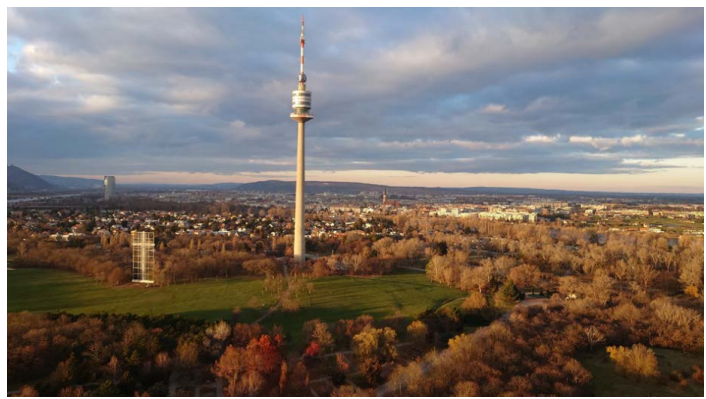

**Figure 15.** View from HJH's office in 25th floor at UN Vienna International Centre across the Danube Park and river, 2023.

05_Sri Lanka_1995 (**Figure 16**)

Report on the Fifth United Nations/European Space Agency Workshop on Basic space Science: From Small Telescopes to Space Missions, hosted by the Arthur C. Clarke Centre for Modern Technologies on behalf of the Government of Sri Lanka (Colombo 11-14 January 1996).

https://www.unoosa.org/oosa/oosadoc/data/documents/1996/aac.105/aac.105640_0.html

https://researchfeatures.com/unbssi-planting-seeds-space-exploration/

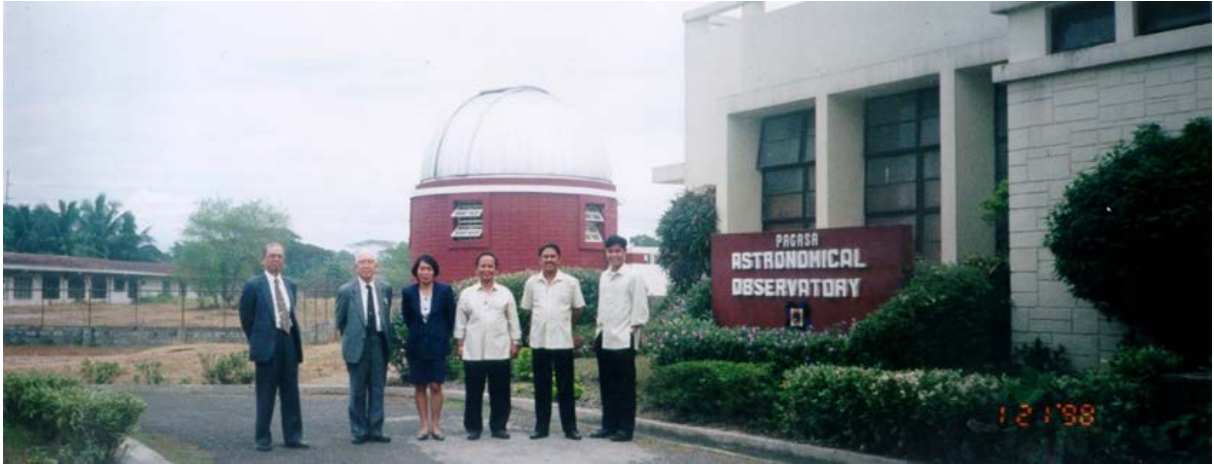

**Figure 16.** T. Kogure, M. Kitamura, and astronomers from the Philippines in front of the observatory housing a telescope from Japan.

06_Germany_1996

Report on the Sixth United Nations/European Space Agency Workshop on Basic Space Science: Ground-Based and Space-Borne Astronomy, hosted by the German Space Agency, on behalf of the Government of Germany, at the Max-Planck-Institute for Radioastronomy (Bonn, Germany, 9-13 September 1996).

https://www.unoosa.org/oosa/oosadoc/data/documents/1996/aac.105/aac.105657_0.html

Proceedings:

https://link.springer.com/journal/10509/volumes-and-issues/258-1

07_Honduras_1997 (**Figure 17**)

Report on the Seventh United Nations/European Space Agency Workshop on Basic Space Science: Small Astronomical Telescopes and Satellites in Education and Research, hosted by the Observatorio Astronómico de la Universidad Nacional Autónoma de Honduras, on behalf of the Government of Honduras (Tegucigalpa, 16-20 June 1997)

https://www.unoosa.org/oosa/oosadoc/data/documents/1997/aac.105/aac.105682_0.html

https://researchfeatures.com/avoiding-armageddon-urgent-search-near-earth-objects/

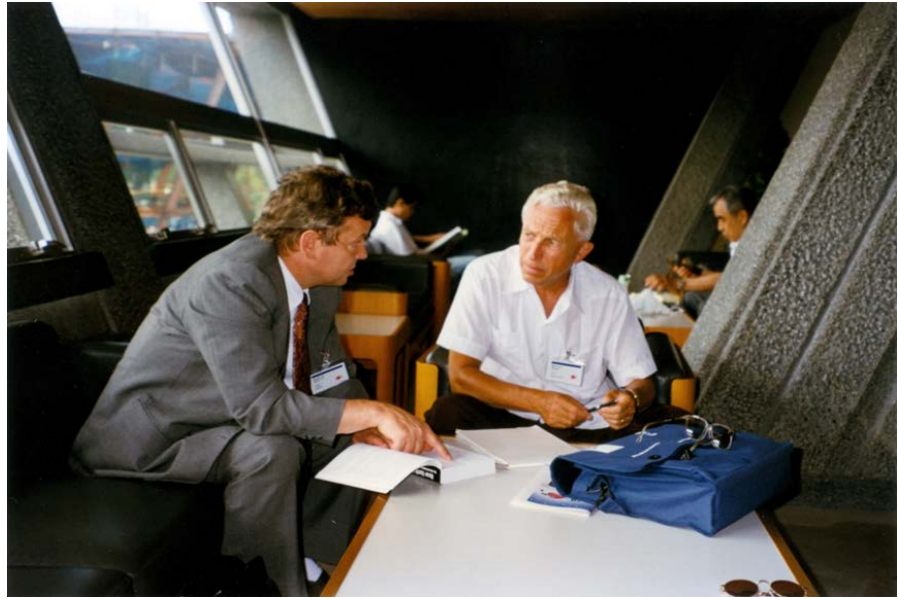

**Figure 17.** HJH meeting eminent university educator D.G. Wentzel (USA) during the IAU meeting in Japan 1997 for getting professional advice for developing astronomy programs in nations around the world.

08_Jordan_1999

Report on the Eighth United Nations/European Space Agency Workshop on Basic Space Science: Scientific Exploration from Space, hosted by the Institute of Astronomy and Space Sciences at Al Al-Bayt University on behalf of the Government of the Hashemite Kingdom of Jordan (Mafraq, Jordan, 13-17 March 1999)

https://www.unoosa.org/oosa/oosadoc/data/documents/1999/aac.105/aac.105723_0.html

Proceedings:

https://link.springer.com/journal/10509/volumes-and-issues/273-1

09_France_2000

Report on the Ninth United Nations/European Space Agency Workshop on Basic Space Science: Satellites and Networks of Telescopes-Tools for Global Participation in the Study of the Universe (Toulouse, France, 27-30 June 2000)

https://www.unoosa.org/oosa/oosadoc/data/documents/2000/aac.105/aac.105742_0.html

10_Mauritius_2001 (**Figure 18**)

Report on the Tenth United Nations/European Space Agency Workshop on Basic Space Science: Exploring the Universe; Sky Surveys, Space Exploration and Space Technologies (Reduit, Mauritius, 25-29 June 2001)

https://www.unoosa.org/oosa/oosadoc/data/documents/2001/aac.105/aac.105766_0.html

Proceedings:

https://link.springer.com/journal/10509/volumes-and-issues/282-1?page=1

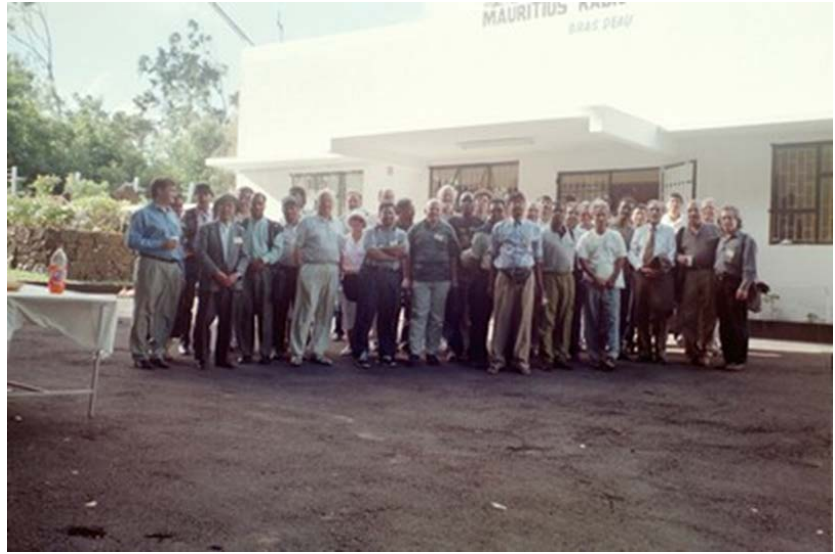

Figure 18. Group photo of the 2001 workshop in Mauritius.

11_Argentina_2002 (**Figure 19** and **Figure 20**)

Report on the Eleventh United Nations/European Space Agency Workshop on Basic Space Science: The World Space Observatory and the Virtual Observatories in the Era of 10-metre Telescopes (Córdoba, Argentina, 9-13 September 2002)

https://www.unoosa.org/oosa/oosadoc/data/docu-ments/2002/aac.105/aac.105784_0.html

https://researchfeatures.com/spektr-uf-unlocking-secrets-uv/

Proceedings:

https://link.springer.com/journal/10509/volumes-and-issues/290-3

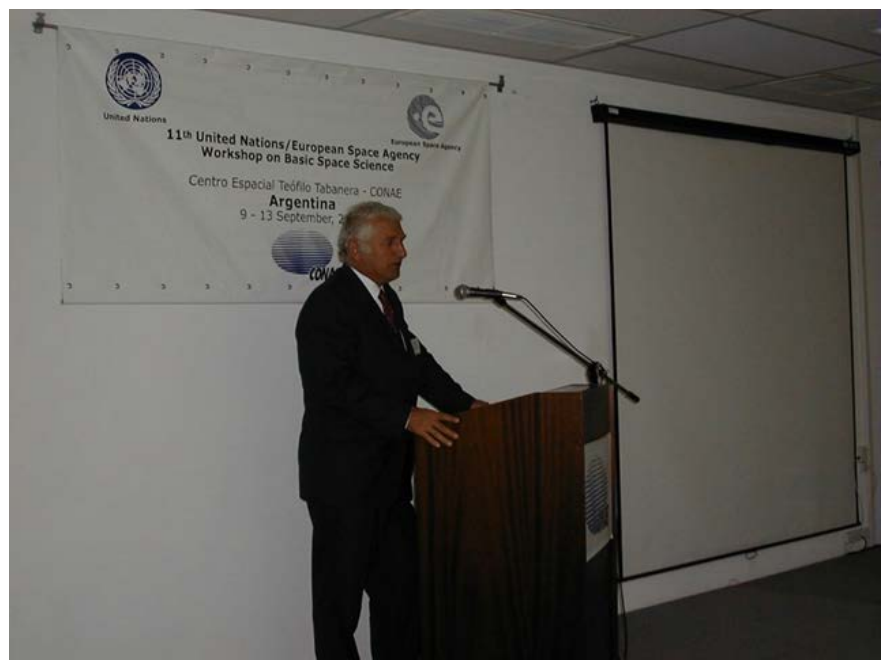

Figure 19. Former chief scientist of the IUE satellite mission Willem Wamsteker (ESA).

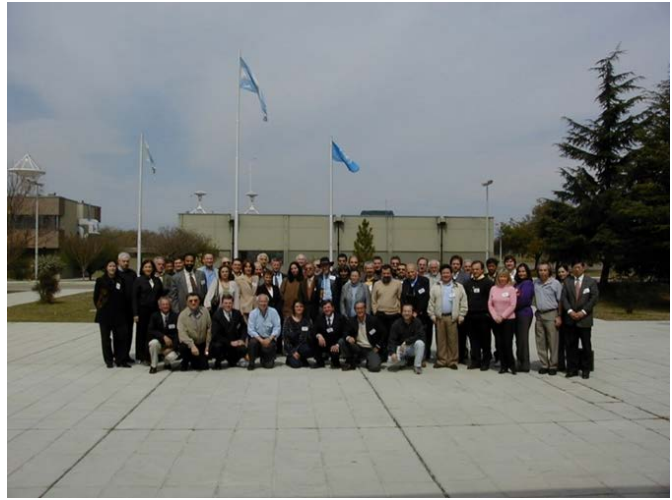

**Figure 20.** Group photo of the 2002 workshop in Argentina.

12_China_2004

Report on the Twelfth United Nations/European Space Agency Workshop on Basic Space Science (Beijing, 24-28 May 2004)

https://www.unoosa.org/oosa/oosadoc/data/documents/2004/aac.105/aac.105829_0.html

Proceedings:

https://link.springer.com/journal/10509/volumes-and-issues/305-3

Five Workshops **International Heliophysical Year 2007 (IHY2007)** (Thompson et al., 2009)

13_United Arab Emirates_2005 (**Figure 21** and **Figure 22**)

Report on the United Nations/European Space Agency/National Aeronautics and Space Administration of the United States of America Workshop on the International Heliophysical Year 2007 (Abu Dhabi and Al-Ain, United Arab Emirates, 20-23 November 2005)

https://www.unoosa.org/oosa/oosadoc/data/documents/2005/aac.105/aac.105856_0.html

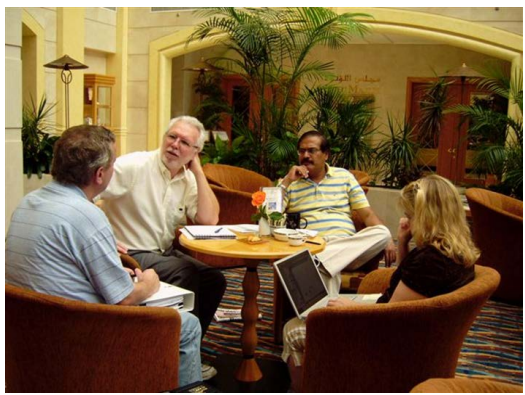

**Figure 21.** HJH, J. Davila (NASA), N. Gopalswami (NASA), and B. Thompson (NASA) meeting for discussions during the 2004 United Arab Emirates workshop.

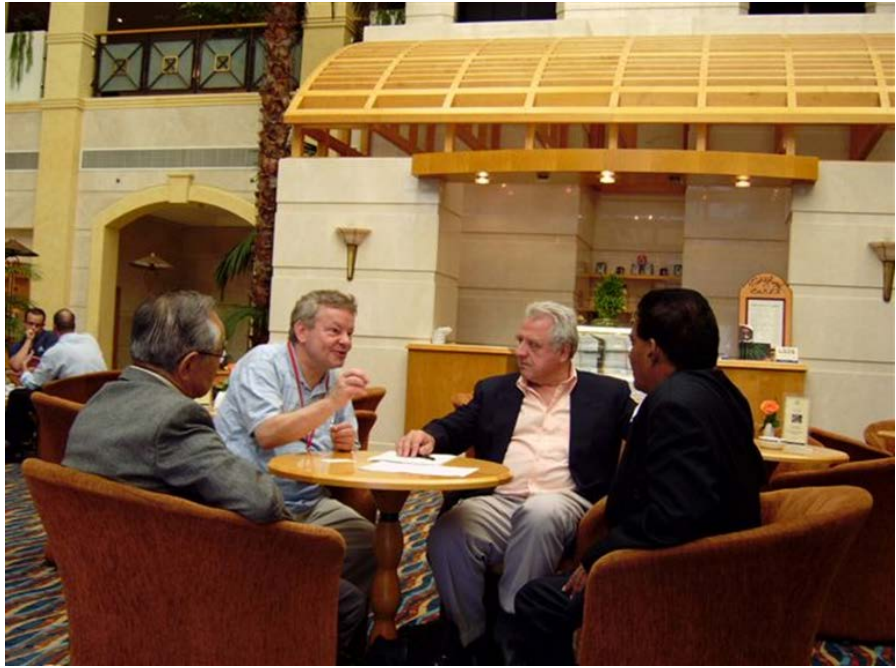

K. Sakurai, HJH, C. Tsallis, and A.M. Mathai discussing the status of the solar neutrino problem during the 2005 workshop in the United Arab Emirates.

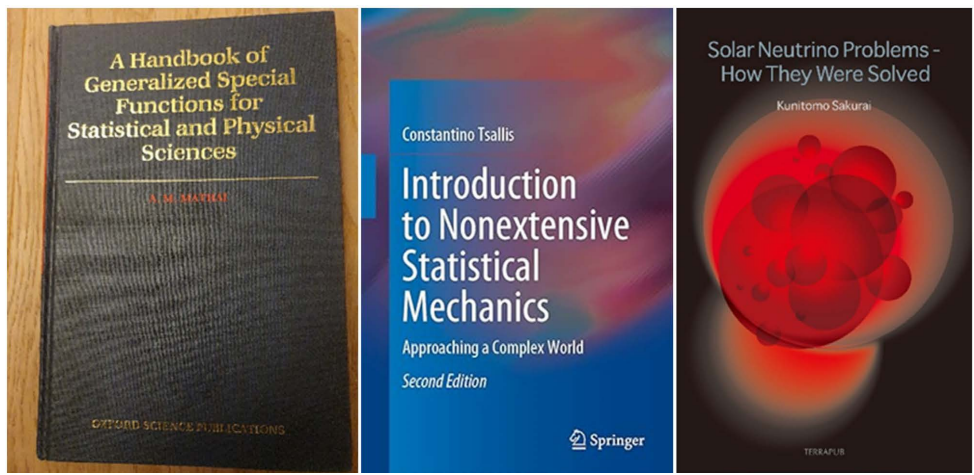

**Figure 22.** Three principal contributors to the workshops from the very beginning in 1991: Mathai (mathematics), Tsallis (physics), Sakurai (astronomy).

14_India_2006 (**Figure 23** and **Figure 24**)

Report on the Second United Nations/National Aeronautics and Space Administration Workshop on the International Heliophysical Year 2007 and Basic Space Science (Bangalore, India, 27 November-1 December 2006).

https://www.unoosa.org/oosa/oosadoc/data/documents/2007/aac.105/aac.105882_0.html

Proceedings:

https://typeset.io/journals/bulletin-of-the-astronomical-society-of-india-bawvm3zu/2007

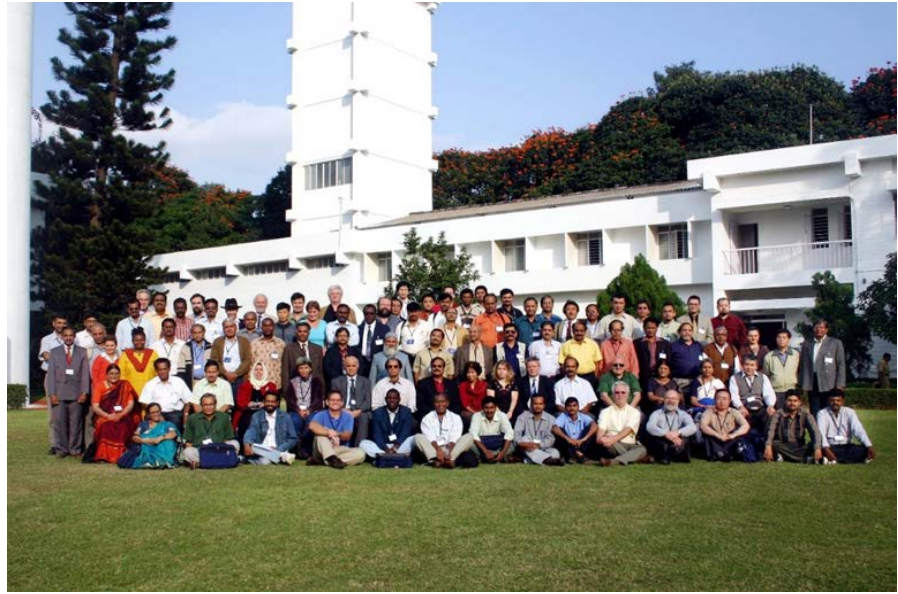

**Figure 23.** Group photo of the 2006 workshop in India.

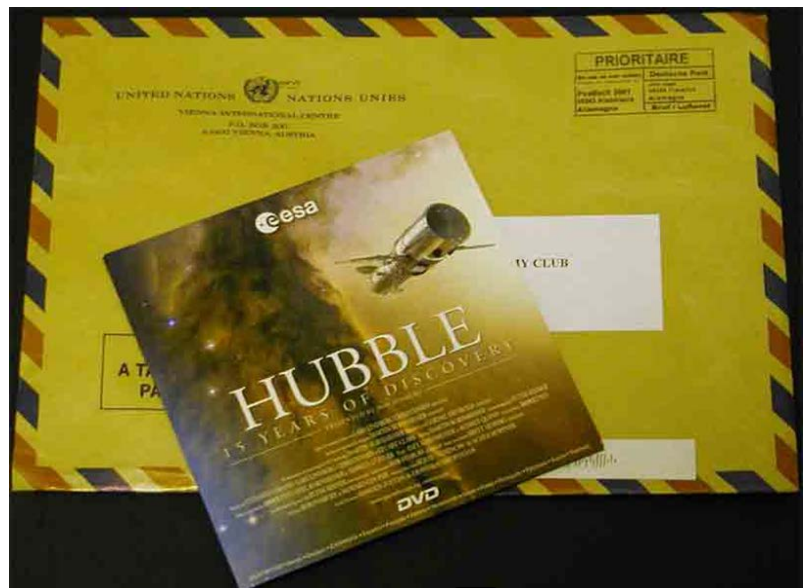

**Figure 24.** HUBBLE: 15 Years of Discovery - DVD.
https://astromart.com/reviews/accessories/software/show/hubble-15-years-of-discovery-dvd

15_Japan_2007 (**Figure 16**, **Figure 17**, **Figure 25** and **Figure 28**)

Report on the Third United Nations/European Space Agency/National Aeronautics and Space Administration Workshop on the International Heliophysical Year 2007 and Basic Space Science (Tokyo, 18-22 June 2007)

https://www.unoosa.org/oosa/oosadoc/data/documents/2007/aac.105/aac.105902_0.html

https://researchfeatures.com/international-heliophysical-year-2007-building-foundation-broader-growth-science/

Proceedings: https://link.springer.com/book/10.1007/978-3-642-03325-4

Proceedings:

https://link.springer.com/journal/11038/volumes-and-issues/104-1

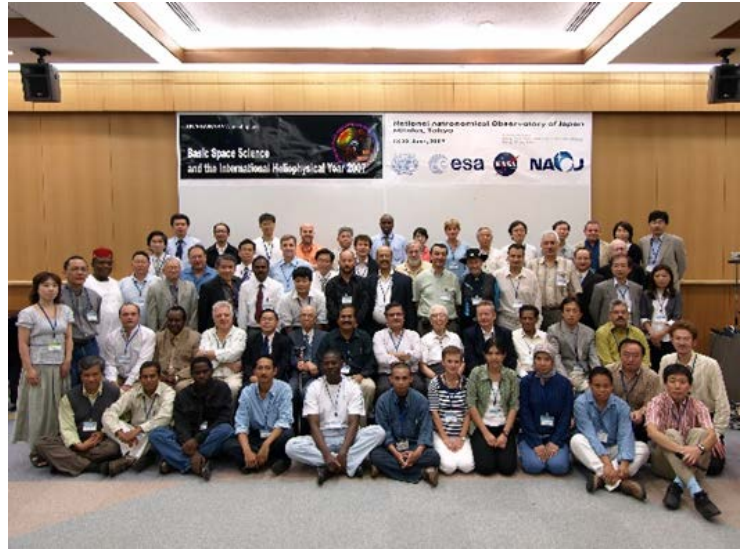

**Figure 25.** Group photo of participants of the 2007 workshop in Japan.

16_Bulgaria_2008 (**Figure 26**)

Report on the Fourth United Nations/ European Space Agency/ National Aeronautics and Space Administration/ Japan Aerospace Exploration Agency Workshop on the International Heliophysical Year 2007 and Basic Space Science (Sozopol, Bulgaria, 2-6 June 2008)

https://www.unoosa.org/oosa/oosadoc/data/documents/2008/aac.105/aac.105919_0.html

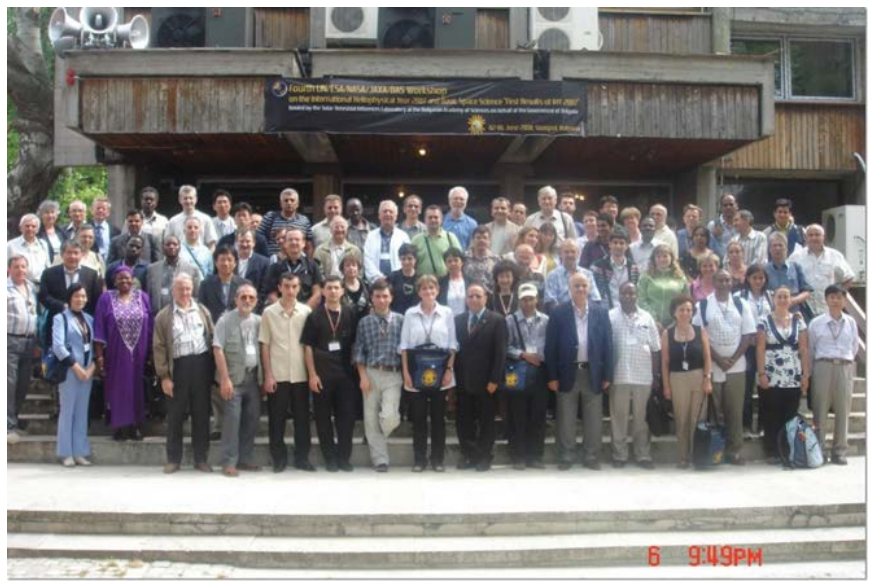

**Figure 26.** Group photo of the 2008 workshop in Bulgaria.

H. J. Haubold, A. M. Mathai

17_Republic of Korea_2009 (**Figure 27**)

Report on the Fifth United Nations/European Space Agency/National Aeronautics and Space Administration/ Japan Aerospace Exploration Agency Workshop on Basic Space Science and the International Heliophysical Year 2007 (Daejeon, Republic of Korea, 21-25 September 2009)

https://www.unoosa.org/oosa/oosadoc/data/documents/2010/aac.105/aac.105964_0.html

Three Workshops **International Space Weather Initiative (ISWI) (Foullon & Malandraki, 2018)**

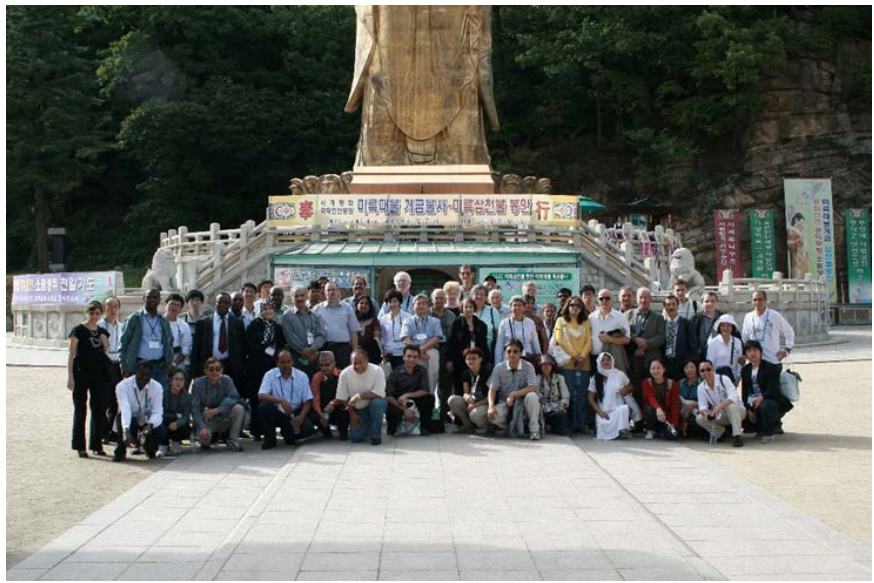

**Figure 27.** Group photo of the 2009 workshop in the Republic of Korea.

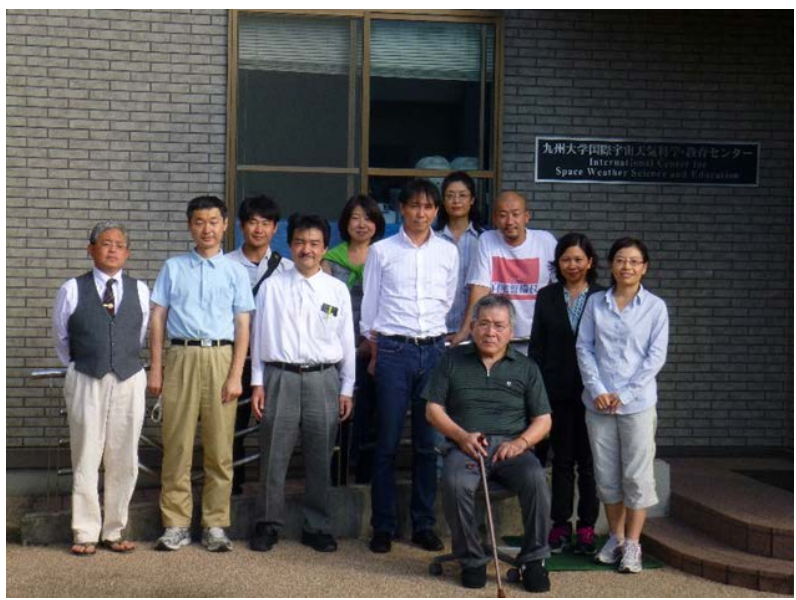

**Figure 28.** K. Yumoto with his professional colleagues in front of the Space Weather Institute Japan.

Eventually, the international activities of IHY 2007 concluded that space weather is the most important issue for global future developments on planet Earth. This led to the holding of three annual space weather workshops 2010-2012.

18_Egypt_2010

Report on the United Nations/National Aeronautics and Space Administration/Japan Aerospace Exploration Agency Workshop on the International Space Weather Initiative (Cairo, 6-10 November 2010)

https://www.unoosa.org/oosa/oosadoc/data/documents/2011/aac.105/aac.105994_0.html

19_Nigeria_2011 (**Figure 29**)

Report on the United Nations/Nigeria Workshop on the International Space Weather Initiative (Abuja, 17-21 October 2011)

https://www.unoosa.org/oosa/oosadoc/data/documents/2012/aac.105/aac.1051018_0.html

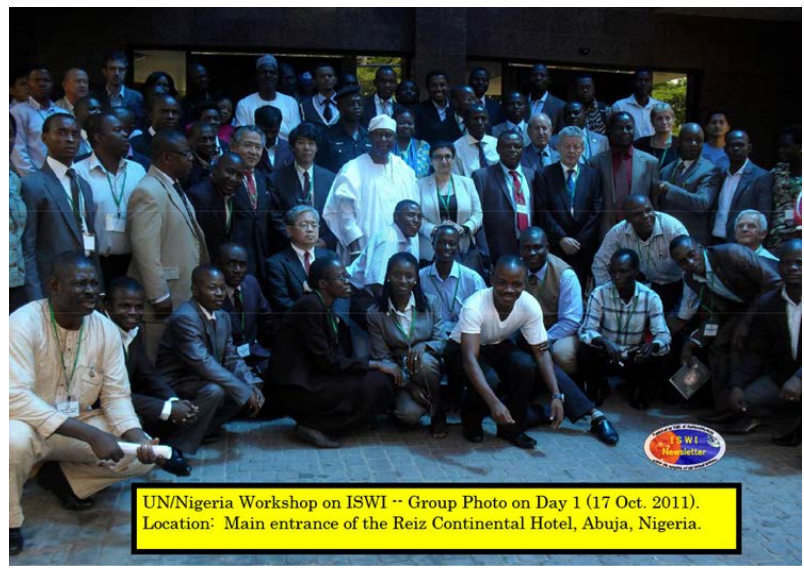

**Figure 29.** Group photo of the 2001 workshop in Nigeria.

20_Ecuador_2012

Report on the United Nations/Ecuador Workshop on the International Space Weather Initiative.

(Quito, 8-12 October 2012)

https://www.unoosa.org/oosa/oosadoc/data/documents/2012/aac.105/aac.1051030_0.html

https://researchfeatures.com/remarkable-legacy-international-space-weather-initiative/

Proceedings: SUN and GEOSPHERE Vol.9, No.1-2014

## 2. Conclusion

The paper serves as a comprehensive guide reporting on educational, teaching,

and research activities in space science and technology over 50 years (1974-2024). It covers various aspects of astronomy, physics, and mathematics and offers access to documents and proceedings in six UN official languages. This work is a result of collaboration with international space agencies and the UN.

The paper provides a holistic overview of the historical development and contributions to space science education and research over a 50-year span, a comprehensive timeline rarely covered in such detail. It documents significant international collaborations and the establishment of educational hubs, presenting an unprecedented multi-agency effort in the field.

Space science and technology is an educational, teaching, and research field that primary focus lies in imparting knowledge about the principles of astronomy, physics, and mathematics that deals with the physical properties and phenomena of celestial bodies and the Universe as a whole. This role requires a deep understanding of complex scientific concepts, astronomy, physics, and mathematics. A space scientist often possesses robust academic qualifications, frequently holding advanced degrees such as a Ph.D. in astrophysics or related scientific disciplines.

In an educational setting, the space scientist plays a critical role in developing educational curricula that articulate the complexities of subjects such as remote sensing, satellite meteorology, satellite communications, basic space science, or satellite navigation, making them accessible to students at various levels of expertise. This involves not only delivering lectures and conducting laboratory sessions but also fostering analytical and critical thinking skills through problem-solving exercises and research projects. Beyond the lecture hall, these space scientists are instrumental in mentoring students, guiding them in pursuing further education or careers in scientific research.

A space scientist plays a crucial role in unravelling the complexities of the Universe for students, blending the wonders of the cosmos with the principles of astronomy, physics, and mathematics. They are responsible for designing and delivering engaging curricula that cover topics ranging from the life cycles of celestial bodies, stars, galaxies, and the Universe. By simplifying advanced theoretical concepts, they help students appreciate the fundamental forces shaping the Universe and develop an understanding of cosmology, stellar evolution, and planetary science. Their methods may include lectures, laboratory experiments, and the use of cutting-edge technology such as telescopes, satellites and computer simulations to provide hands-on learning experiences. Space scientists also guide students in scientific inquiry, encouraging them to pose meaningful questions and pursue research projects that challenge existing paradigms. Furthermore, they stay up to date with the latest developments in the field, often engaging in ongoing research and contributing to scientific publications. Art and science are widely seen as being separate from each other, but according to Hans and Barbara Haubold at the Vienna International Centre, Austria, that is far from the case. Hans and Barbara explore how a group of 19th century American landscape artists was inspired by

H. J. Haubold, A. M. Mathaithe rapid pace of scientific progress at the time and how their changing motives were reflected in scientific debates. They also show how the latest advances in technology could even help to deepen our understanding and appreciation of artwork and its evolution throughout history.

After 50 years of research, teaching, and education efforts in space science and technology, supported by the United Nations, we can conclude that research has contributed to the solution of the solar neutrino problem, teaching is operational in the UN-affiliated regional centres for space science and technology, and we follow with interest the establishment of education at the Albert Einstein Discovery Center Ulm e.V. (Germany, https://einstein.center).

Despite their enormous abundance in the universe, neutrinos remain the least understood fundamental particles of the standard model of elementary interactions and elementary particles. Further data acquisition in neutrino experiments may lead to a better understanding of the paradox of entropy production in anomalous diffusion and the related switch between quantum mechanics and classical mechanics using fractional calculus (**Figure 30**).

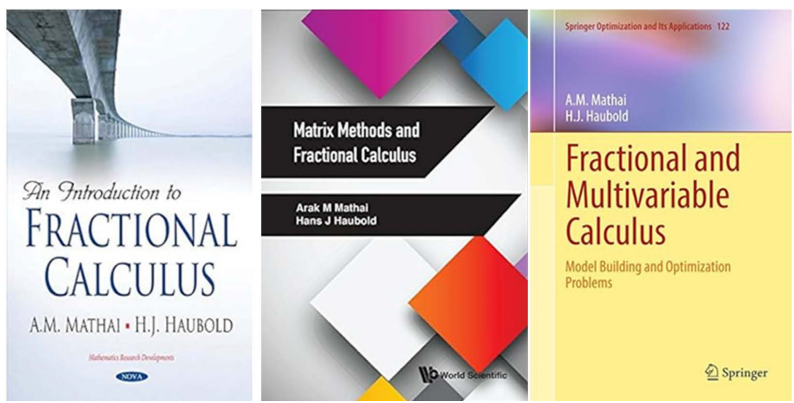

Three monographs being used to lecture on fractional calculus and applications.

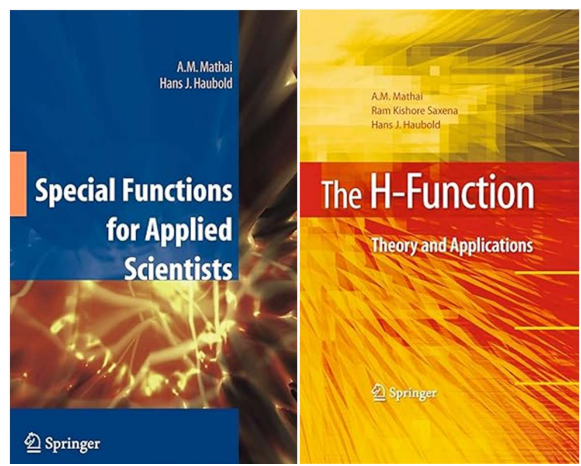

**Figure 30.** Fractional reaction and diffusion are two topics in classical and quantum statistical mechanics that invite the use of special functions of mathematical physics for closed form evaluation of respective differential equations.

DOI: 10.4236/***.2025.***** 27 Creative Education

## Conflicts of Interest

The authors declare no conflicts of interest regarding the publication of this paper.

## References

Alpher, R. A., & Herman, R. (2001). *Genesis of the Big Bang*. Oxford University Press.

Capacity Building (2008). *Capacity-Building in Space Science and Technology, Regional Centres for Space Science and Technology Education Affiliated to the United Nations.* United Nations, Office for Outer Space Affairs.
https://www.unoosa.org/pdf/publications/st_space_41E.pdf

Foullon, C., & Malandraki, O. E. (2018). *Space Weather of the Heliosphere: Processes and Forecasts.* Cambridge University Press.

Gottloeber, S., Haubold, H. J., Liebscher, D. E., Muecket, J. P., & Mueller, V. (1990). Relativistic Astrophysics and Gravitation. *Astronomische Nachrichten, 311,* 145-145.
https://doi.org/10.1002/asna.2113110302

Haubold, H. J. (2020). A. M. Mathai Centre for Mathematical and Statistical Sciences: A Brief History of the Centre and Prof. Dr. A. M. Mathai's Research and Education Programs at the Occasion of His 85th Anniversary. *Creative Education, 11,* 356-405.
https://doi.org/10.4236/ce.2020.113028

Haubold, H. J. (2024). History and Education of the Albert A. Michelson Exhibition Developed at the Occasion of the Einstein Centenary Berlin 1979 and the Michelson Colloquium Potsdam 1981. *Creative Education, 15,* 1166-1194.
https://doi.org/10.4236/ce.2024.156071

Haubold, H. J., & Mathai, A. M. (1997). Sun. In J. H. Shirley, & R. W. Fairbridge (Eds.), *Encyclopedia of Planetary Sciences* (pp. 786-794). Kluwer Academic Publishers.
https://doi.org/10.1007/1-4020-4520-4_390

Haubold, H. J., & Mathai, A. M. (1998) Structure of the Universe. In G. L. Trigg (Ed.), *Encyclopedia of Applied Physics: Vol. 23* (pp. 47-82). Wiley.

Mathai, A. M. (2011). *Centre for Mathematical Sciences: Profile 1977-2011.*
https://en.wikipedia.org/wiki/Centre_for_Mathematical_Sciences_(Kerala)

Mathai, A. M., & Haubold, H. J. (2018). United Nations Basic Space Science Initiative (UNBSSI) 1991-2012 and beyond. *Creative Education, 9,* 192-248.
https://doi.org/10.4236/ce.2018.92015

Michelson Livingston, D. (1973). *The Master of Light: A Biography of Albert A. Michelson.* Charles Scribner's Sons.

Nernst, W. (1914). *Die Theorie der Strahlung und der Quanten, Verhandlungen auf einer von E. Solvay einberufenen Zusammenkunft (30, Oktober bis 3. November 1911), mit einem Anhange über die Entwicklung der Quantentheorie vom Herbst 1911 bis zum Sommer 1913, in deutscher Sprache herausgegeben von A. Eucken.* von Wilhelm Knapp.

Pyenson, L., Mathai, A. M., & Haubold, H. J. (2019). United Nations Education Program in Space Science and Technology 1988-2018. *Creative Education, 10,* 2219-2231.
https://doi.org/10.4236/ce.2019.1010160

Reines, F. (1972). *Cosmology, Fusion & Other Matters, George Gamow Memorial Volume.* Colorado Associated University Press.

Strauch, D. (2015). *Einsteins Sommer-Idyll in Caputh: Biographie eines Sommerhauses.* Edition progris.

Thompson, B. J., Gopalswamy, N., Davila, J. M., & Haubold, H. J. (2009). *Putting the "I"*

in IHY: The United Nations Report for the International Heliophysical Year 2007. Springer.

Treder, H. J., & Rompe. R. (1982). Albert. A. Michelson Colloquium. *Astronomische Nachrichten, 303,* 91-96.

Wamsteker, W., Albrecht, R., & Haubold, H. J. (2004). *Developing Basic Space Sciences World-Wide: A Decade of UN/ESA Workshops.* Kluwer Academic Publishers.